\documentclass[final]{cvpr}

\usepackage{times}
\usepackage{epsfig}
\usepackage{graphicx}
\usepackage{amsmath}
\usepackage{amssymb}
\usepackage{dblfloatfix}
\usepackage[table,x11names]{xcolor}

\usepackage[pagebackref=true,breaklinks=true,colorlinks,bookmarks=false]{hyperref}

\def\cvprPaperID{****} 
\def\confYear{CVPR 2021}
\definecolor{redhighlight}{RGB}{255,247,247}

\newsavebox\CBox
\def\textBF#1{\sbox\CBox{#1}\resizebox{\wd\CBox}{\ht\CBox}{\textbf{#1}}}

\begin{document}

\title{
\vspace{-8mm}
Fast and Accurate Single-Image Depth Estimation on Mobile Devices,\\ Mobile AI 2021 Challenge: Report}
\author{
Andrey Ignatov \and Grigory Malivenko \and David Plowman \and Samarth Shukla \and Radu Timofte \and
Ziyu Zhang \and Yicheng Wang \and Zilong Huang \and Guozhong Luo \and Gang Yu \and Bin Fu \and
Yiran Wang \and  Xingyi Li \and  Min Shi \and Ke Xian \and  Zhiguo Cao \and
Jin-Hua Du \and Pei-Lin Wu \and Chao Ge \and
Jiaoyang Yao \and Fangwen Tu \and Bo Li \and
Jung Eun Yoo  \and Kwanggyoon Seo  \and
Jialei Xu \and Zhenyu Li \and Xianming Liu \and Junjun Jiang \and
Wei-Chi Chen \and
Shayan Joya \and
Huanhuan Fan \and Zhaobing Kang \and Ang Li \and Tianpeng Feng \and
Yang Liu \and Chuannan Sheng \and Jian Yin \and
Fausto T. Benavides
}

\maketitle

\maketitle

\begin{abstract}

Depth estimation is an important computer vision problem with many practical applications to mobile devices. While many solutions have been proposed for this task, they are usually very computationally expensive and thus are not applicable for on-device inference. To address this problem, we introduce the first Mobile AI challenge, where the target is to develop an end-to-end deep learning-based depth estimation solutions that can demonstrate a nearly real-time performance on smartphones and IoT platforms. For this, the participants were provided with a new large-scale dataset containing RGB-depth image pairs obtained with a dedicated stereo ZED camera producing high-resolution depth maps for objects located at up to 50 meters. The runtime of all models was evaluated on the popular Raspberry Pi 4 platform with a mobile ARM-based Broadcom chipset. The proposed solutions can generate VGA resolution depth maps at up to 10 FPS on the Raspberry Pi 4 while achieving high fidelity results, and are compatible with any Android or Linux-based mobile devices. A detailed description of all models developed in the challenge is provided in this paper.

\end{abstract}
{\let\thefootnote\relax\footnotetext{%
\hspace{-5mm}$^*$
Andrey Ignatov, Grigory Malivenko, David Plowman and Radu Timofte are the Mobile AI 2021 challenge organizers \textit{(andrey@vision.ee.ethz.ch, grigory.malivenko@gmail.com, david.plowman@raspberrypi.com, radu.timofte@vision.ee.ethz.ch)}. The other authors participated in the challenge. Appendix \ref{sec:apd:team} contains the authors' team names and affiliations. \vspace{2mm} \\ Mobile AI 2021 Workshop website: \\ \url{https://ai-benchmark.com/workshops/mai/2021/}
}}

\vspace{-2mm}
\section{Introduction}

\begin{figure*}[t!]
\centering
\setlength{\tabcolsep}{1pt}
\resizebox{\linewidth}{!}
{
\includegraphics[width=0.5\linewidth]{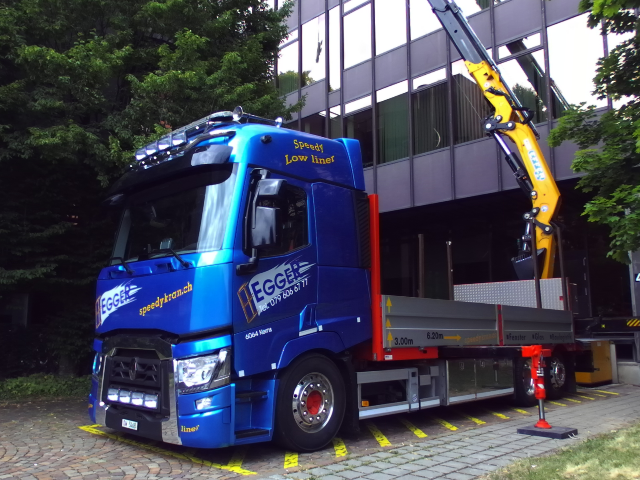} \hspace{2mm}
\includegraphics[width=0.5\linewidth]{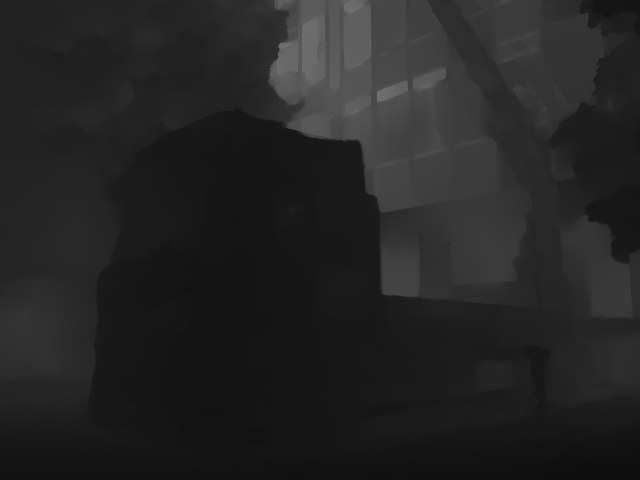}
}
\vspace{-1.2mm}
\caption{The original RGB image and the corresponding depth map obtained with the ZED 3D camera.}
\label{fig:example_photos}
\vspace{-1.2mm}
\end{figure*}

A wide spread of various depth-guided problems related to augmented reality, gesture recognition, object segmentation, autonomous driving and bokeh effect rendering tasks has created a strong demand for fast and efficient single-image depth estimation approaches that can run on portable low-power hardware. While many accurate deep learning-based solutions have been proposed for this problem in the past~\cite{li2018megadepth,godard2019digging,eigen2014depth,liu2015deep,liu2015learning,laina2016deeper,garg2016unsupervised,chen2016single}, they were optimized for high fidelity results only while not taking into account computational efficiency and mobile-related constraints, which is essential for tasks related to image processing~\cite{ignatov2017dslr,ignatov2018wespe,ignatov2020replacing} on mobile devices. This results in solutions requiring powerful high-end GPUs and consuming gigabytes of RAM when processing even low-resolution input data, thus being incompatible with resource-constrained mobile hardware. In this challenge, we change the current depth estimation benchmarking paradigm by using a new depth estimation dataset collected in the wild and by imposing additional efficiency-related constraints on the designed solutions.

When it comes to the deployment of AI-based solutions on portable devices, one needs to take care of the particularities of mobile CPUs, NPUs and GPUs to design an efficient model. An extensive overview of mobile AI acceleration hardware and its performance is provided in~\cite{ignatov2019ai,ignatov2018ai}. According to the results reported in these papers, the latest mobile NPUs are already approaching the results of mid-range desktop GPUs released not long ago. However, there are still two major issues that prevent a straightforward deployment of neural networks on mobile devices: a restricted amount of RAM, and a limited and not always efficient support for many common deep learning layers and operators. These two problems make it impossible to process high resolution data with standard NN models, thus requiring a careful adaptation of each architecture to the restrictions of mobile AI hardware. Such optimizations can include network pruning and compression~\cite{chiang2020deploying,ignatov2020rendering,li2019learning,liu2019metapruning,obukhov2020t}, 16-bit / 8-bit~\cite{chiang2020deploying,jain2019trained,jacob2018quantization,yang2019quantization} and low-bit~\cite{cai2020zeroq,uhlich2019mixed,ignatov2020controlling,liu2018bi} quantization, device- or NPU-specific adaptations, platform-aware neural architecture search~\cite{howard2019searching,tan2019mnasnet,wu2019fbnet,wan2020fbnetv2}, \etc.

While many challenges and works targeted at efficient deep learning models have been proposed recently, the evaluation of the obtained solutions is generally performed on desktop CPUs and GPUs, making the developed solutions not practical due to the above mentioned issues. To address this problem, we introduce the first \textit{Mobile AI Workshop and Challenges}, where all deep learning solutions are developed for and evaluated on real low-power devices.
In this competition, the participating teams were provided with a novel depth estimation dataset containing over 8 thousand RGB-depth image pairs collected in the wild with a stereo ZED 3D camera. Within the challenge, the participants were evaluating the runtime and tuning their models on the Raspberry Pi 4 ARM based single-board computer used as a target platform for many embedded machine learning projects.
The final score of each submitted solution was based on the runtime and fidelity results, thus balancing between the image reconstruction quality and efficiency of the proposed model. Finally, all developed solutions are fully compatible with the TensorFlow Lite framework~\cite{TensorFlowLite2021}, thus can be deployed and accelerated on any mobile platform providing AI acceleration through the Android Neural Networks API (NNAPI)~\cite{NNAPI2021} or custom TFLite delegates~\cite{TFLiteDelegates2021}.

\smallskip


This challenge is a part of the \textit{MAI 2021 Workshop and Challenges} consisting of the following competitions:


\small

\begin{itemize}
\item Learned Smartphone ISP on Mobile NPUs~\cite{ignatov2021learned}
\item Real Image Denoising on Mobile GPUs~\cite{ignatov2021fastDenoising}
\item Quantized Image Super-Resolution on Edge SoC NPUs~\cite{ignatov2021real}
\item Real-Time Video Super-Resolution on Mobile GPUs~\cite{romero2021real}
\item Single-Image Depth Estimation on Mobile Devices
\item Quantized Camera Scene Detection on Smartphones~\cite{ignatov2021fastSceneDetection}
\item High Dynamic Range Image Processing on Mobile NPUs
\end{itemize}

\normalsize


\noindent The results obtained in the other competitions and the description of the proposed solutions can be found in the corresponding challenge papers.


\begin{figure*}[t!]
\centering
\setlength{\tabcolsep}{1pt}
\resizebox{0.96\linewidth}{!}
{
\includegraphics[width=1.0\linewidth]{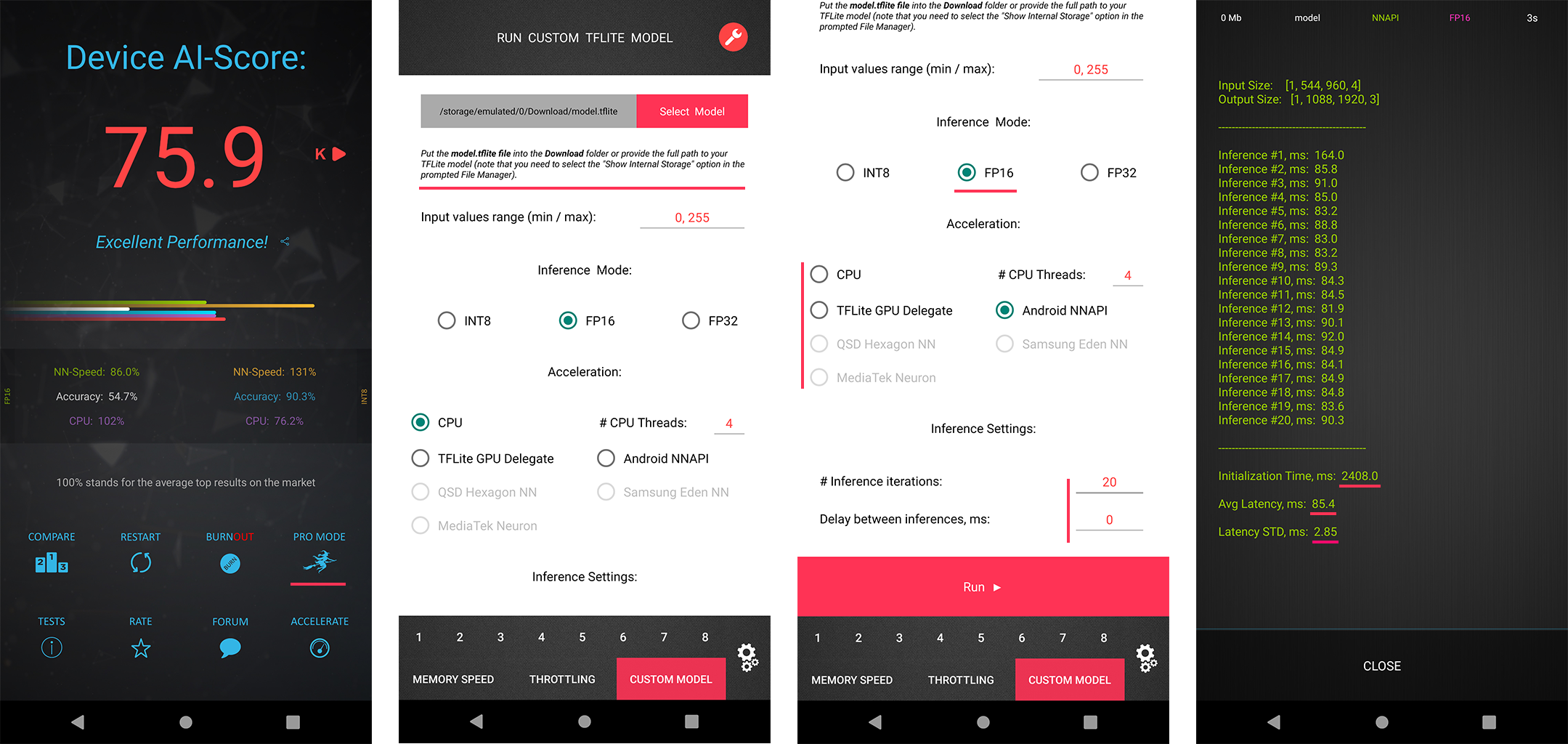}
}
\vspace{0.2cm}
\caption{Loading and running custom TensorFlow Lite models with AI Benchmark application. The currently supported acceleration options include Android NNAPI, TFLite GPU, Hexagon NN, Samsung Eden and MediaTek Neuron delegates as well as CPU inference through TFLite or XNNPACK backends. The latest app version can be downloaded at \url{https://ai-benchmark.com/download}}
\vspace{-1.2mm}
\label{fig:ai_benchmark_custom}
\end{figure*}

\section{Challenge}

To develop an efficient and practical solution for mobile-related tasks, one needs the following major components:

\begin{enumerate}
\item A high-quality and large-scale dataset that can be used to train and evaluate the solution;
\item An efficient way to check the runtime and debug the model locally without any constraints;
\item An ability to regularly test the runtime of the designed neural network on the target mobile platform or device.
\end{enumerate}

This challenge addresses all the above issues. Real training data, tools, and runtime evaluation options provided to the challenge participants are described in the next sections.

\subsection{Dataset}

To get real and diverse data for the considered challenge, a novel dataset consisting of RGB-depth image pairs was collected using the ZED stereo camera\footnote{\url{https://www.stereolabs.com/zed/}} capable of shooting 2K images. It demonstrates an average depth estimation error of less than 0.2m for objects located closer than 8 meters~\cite{ortiz2018depth}, while more coarse predictions are also available for distances of up to 50 meters. Around 8.3K image pairs were collected in the wild over several weeks in a variety of places. For this challenge, the obtained images were downscaled to VGA resolution (640$\times$480 pixels) that is typically used on mobile devices for different depth-related tasks. The original RGB images were then considered as inputs, and the corresponding 16-bit depth maps~--- as targets. A sample RGB-depth image pair from the collected dataset is demonstrated in Fig.~\ref{fig:example_photos}.

\subsection{Local Runtime Evaluation}

When developing AI solutions for mobile devices, it is vital to be able to test the designed models and debug all emerging issues locally on available devices. For this, the participants were provided with the \textit{AI Benchmark} application~\cite{ignatov2018ai,ignatov2019ai} that allows to load any custom TensorFlow Lite model and run it on any Android device with all supported acceleration options. This tool contains the latest versions of \textit{Android NNAPI, TFLite GPU, Hexagon NN, Samsung Eden} and \textit{MediaTek Neuron} delegates, therefore supporting all current mobile platforms and providing the users with the ability to execute neural networks on smartphone NPUs, APUs, DSPs, GPUs and CPUs.

\smallskip

To load and run a custom TensorFlow Lite model, one needs to follow the next steps:

\begin{enumerate}
\setlength\itemsep{0mm}
\item Download AI Benchmark from the official website\footnote{\url{https://ai-benchmark.com/download}} or from the Google Play\footnote{\url{https://play.google.com/store/apps/details?id=org.benchmark.demo}} and run its standard tests.
\item After the end of the tests, enter the \textit{PRO Mode} and select the \textit{Custom Model} tab there.
\item Rename the exported TFLite model to \textit{model.tflite} and put it into the \textit{Download} folder of the device.
\item Select mode type \textit{(INT8, FP16, or FP32)}, the desired acceleration/inference options and run the model.
\end{enumerate}

\noindent These steps are also illustrated in Fig.~\ref{fig:ai_benchmark_custom}.

\subsection{Runtime Evaluation on the Target Platform}

In this challenge, we use the \textit{Raspberry Pi 4} single-board computer as our target runtime evaluation platform. It is based on the \textit{Broadcom BCM2711} chipset containing four Cortex-A72 ARM cores clocked at 1.5 GHz and demonstrates AI Benchmark scores comparable to entry-level Android smartphone SoCs~\cite{AIBenchmark202104}. The Raspberry Pi 4 supports the majority of Linux distributions, Windows 10 IoT build as well as Android operating system. In this competition, the runtime of all solutions was tested using the official TensorFlow Lite 2.5.0 Linux build~\cite{TensorFlowLite2021Linux} containing many important performance optimizations for the above chipset, the default \textit{Raspberry Pi OS} was installed on the device. Within the challenge, the participants were able to upload their TFLite models to the runtime validation server connected to a real Raspberry Pi 4 board and get instantaneous feedback: the runtime of their solution or an error log if the model contains some incompatible operations. The same setup was also used for the final runtime evaluation.

\begin{table*}[t!]
\centering
\resizebox{\linewidth}{!}
{
\begin{tabular}{l|c|cc|cccc|c|c}
\hline
Team \, & \, Author \, & \, Framework \, & \, Model Size, MB \, & \, si-RMSE$\downarrow$ \, & \, RMSE$\downarrow$ \, & \, LOG10$\downarrow$ \, & \, REL$\downarrow$ \, & \, Runtime, ms $\downarrow$ \, & \, Final Score \\
\hline
\hline
Tencent GY-Lab & \, Parkzyzhang \, & \, PyTorch / TensorFlow \, &   3.4 & 0.2836 & 3.56 & 0.1121 & 0.2690  &    \textBF{97} & \textBF{129.41} \\
SMART & KX\_SMART         & PyTorch / TensorFlow &  15.0 & 0.2602 & 3.25 & 0.1043 & 0.2678  &  1197 &  14.51 \\
Airia-Team1 & dujinhua    & TensorFlow           &  64.9 & 0.2408 & 3.00 & 0.0904 & 0.2389  &  1933 &  11.75 \\
YTL & Jacob.Yao           & PyTorch / TensorFlow &  56.2 & 0.2902 & 3.91 & 0.1551 & 0.4700  &  1275 &   8.98 \\
CFL2 & jey                & PyTorch / TensorFlow &   9.6 & 0.2761 & 9.68 & 2.3393 & 0.9951  &   772 &   5.5  \\
HIT-AIIA & zhyl           & Keras / TensorFlow   &  56.0 & \textBF{0.2332} & \textBF{2.72} & \textBF{0.0831} & \textBF{0.2189}  &  6146 &   4.11 \\
weichi & weichi           & TensorFlow           &   0.5 & 0.4659 & 7.56 & 0.4493 & 0.5992  &   582 &   1.72 \\
MonoVision Palace \,      & shayanj & TensorFlow &  15.3 & 0.3543 & 4.16 & 0.1441 & 0.3862  &  3466 &   1.36 \\
3dv oppo & fanhuanhuan    & PyTorch / TensorFlow &   187 & 0.2678 & 5.96 & 0.3300 & 0.5152  & 26494 &   0.59 \\
MegaUe & faustChok         & Keras / TensorFlow   &   118 & 0.3737 & 9.08 & 0.9605 & 0.8573  &  9392 &   0.38 \\
\end{tabular}
}
\vspace{2.6mm}
\caption{\small{MAI 2021 Monocular Depth Estimation challenge results and final rankings. The runtime values were obtained on 640$\times$480 px images on the Raspberry Pi 4 device. Team \textit{Tencent GY-Lab} is the challenge winner, the best fidelity results are obtained by team \textit{HIT-AIIA}.}}
\label{tab:results}
\end{table*}

\subsection{Challenge Phases}

The challenge consisted of the following phases:

\vspace{-0.8mm}
\begin{enumerate}
\item[I.] \textit{Development:} the participants get access to the data and AI Benchmark app, and are able to train the models and evaluate their runtime locally;
\item[II.] \textit{Validation:} the participants can upload their models to the remote server to check the fidelity scores on the validation dataset, to get the runtime on the target platform, and to compare their results on the validation leaderboard;
\item[III.] \textit{Testing:} the participants submit their final results, codes, TensorFlow Lite models, and factsheets.
\end{enumerate}
\vspace{-0.8mm}

\subsection{Scoring System}

All solutions were evaluated using the following metrics:

\vspace{-0.8mm}
\begin{itemize}
\setlength\itemsep{-0.2mm}
\item Root Mean Squared Error (RMSE) measuring the absolute depth estimation accuracy,
\item Scale Invariant Root Mean Squared Error (si-RMSE) measuring the quality of relative depth estimation (relative position of the objects),
\item Average $\log_{10}$ and Relative (REL) errors~\cite{liu2015learning},
\item The runtime on the target Raspberry Pi 4 device.
\end{itemize}
\vspace{-0.8mm}

The score of each final submission was evaluated based on the next formula ($C$ is a constant normalization factor):

\smallskip
\begin{equation*}
\text{Final Score} \,=\, \frac{2^{-20 \cdot \text{si-RMSE}}}{C \cdot \text{runtime}},
\end{equation*}
\smallskip

During the final challenge phase, the participants did not have access to the test dataset. Instead, they had to submit their final TensorFlow Lite models that were subsequently used by the challenge organizers to check both the runtime and the fidelity results of each submission under identical conditions. This approach solved all the issues related to model overfitting, reproducibility of the results, and consistency of the obtained runtime/accuracy values.

\begin{table*}[t!]
\centering
\resizebox{\linewidth}{!}
{
\begin{tabular}{l|cc|cc|cc|cc}
\hline
Mobile SoC & \, Snapdragon 888 \, & \,  Snapdragon 855 \,  & \,  Dimensity 1000 \,  & \,  Dimensity 800 \,  & \,  Exynos 2100 \,  & \, Exynos 990 \, & \, Kirin 990 5G \, & \, Kirin 980 \, \\
GPU & \, \small Adreno 660, ms \, & \, \small Adreno 640, ms \, & \, \small Mali-G77 MP9, ms \, & \, \small Mali-G57 MP4, ms \, & \, \small Mali-G78 MP14, ms \, & \, \small Mali-G77 MP11, ms \, & \, \small Mali-G76 MP16, ms \, & \, \small Mali-G76 MP10, ms \, \\
\hline
Tencent GY-Lab          & 3.5  & 5.7  & 8.6  & 13   & 5.7  & 12   & 8.8  & 9.3  \\
SMART                   & 33   & 60   & 65   & 106  & 37   & 53   & 48   & 58   \\
\rowcolor{redhighlight} Airia-Team1 $^*$        & 283  & 321  & 295  & 447  & 248  & 270  & 337  & 351  \\
YTL                     & 35   & 70   & 71   & 104  & 36   & 52   & 54   & 65   \\
\rowcolor{redhighlight} CFL2 $^*$               & 121  & 179  & 186  & 277  & 117  & 170  & 179  & 188  \\
HIT-AIIA                & 95   & 175  & 149  & 320  & 101  & 137  & 142  & 183  \\
weichi                  & 7.1  & 11   & 23   & 43   & 13   & 18   & 18   & 22   \\
MonoVision Palace \,    & 77   & 128  & 119  & 247  & 71   & 97   & 101  & 129  \\
\rowcolor{redhighlight} 3dv oppo $^*$           & 3672 & 4346 & 4053 & 4832 & 4071 & 3649 & 3753 & 4107 \\
MegaUe                   & 141  & 288  & 245  & 547  & 182  & 234  & 209  & 266  \\
\end{tabular}
}
\vspace{2.6mm}
\caption{\small{The speed of the proposed solutions on several popular mobile GPUs. The runtime was measured with the AI Benchmark app using the TFLite GPU delegate~\cite{lee2019device}. $^*$ Solutions from teams \textit{Airia-Team1}, \textit{CFL2} and \textit{3dv oppo} are not compatible with neither TFLite delegates nor Android NNAPI due to the issues related to PyTorch $\rightarrow$ TFLite conversion, thus were executed on mobile CPUs.}}
\label{tab:runtime_results}
\end{table*}

\section{Challenge Results}

From above 140 registered participants, 10 teams entered the final phase and submitted valid results, TFLite models, codes, executables and factsheets. Table~\ref{tab:results} summarizes the final challenge results and reports si-RMSE, RMSE, LOG10 and REL measures and runtime numbers for each submitted solution on the final test dataset and on the target evaluation platform. The proposed methods are described in section~\ref{sec:solutions}, and the team members and affiliations are listed in Appendix~\ref{sec:apd:team}.

\begin{figure*}[b!]
\vspace{2mm}
\centering
\includegraphics[width=0.86\textwidth]{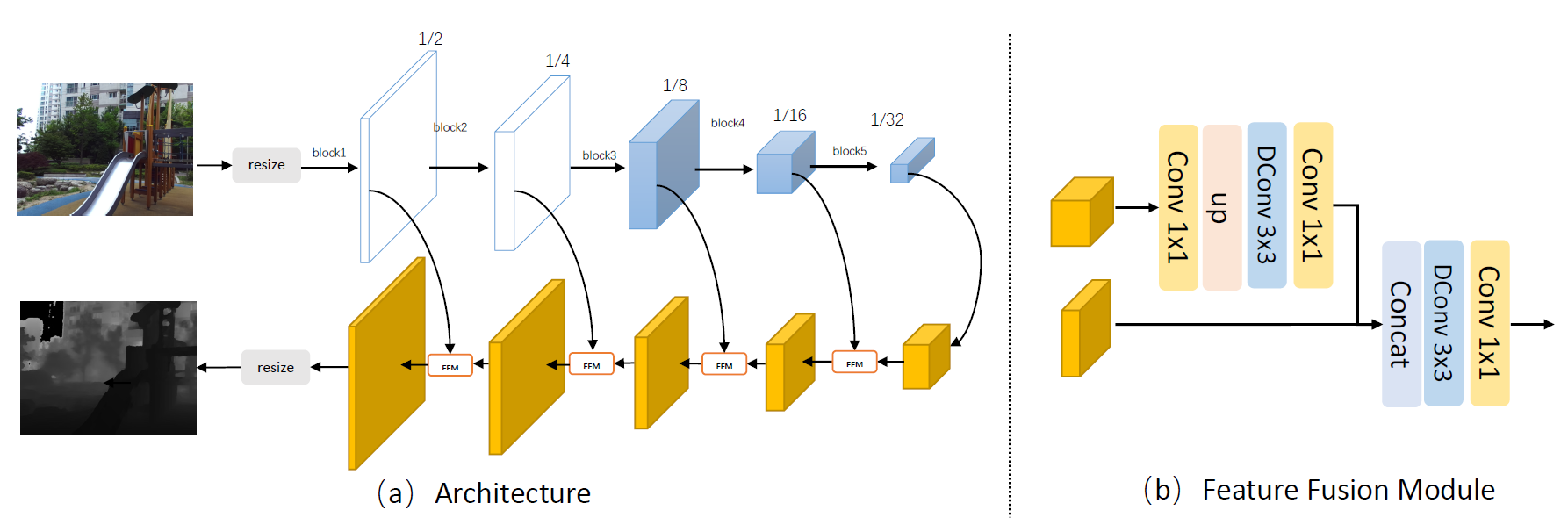}
\vspace{2mm}
\caption{The model architecture and the structure of the Feature Fusion Module (FFM) proposed by team Tencent GY-Lab.}
\label{fig:light_architecture}
\end{figure*}

\subsection{Results and Discussion}

All proposed solutions are relying on the encoder-decoder based architecture as it allows both to perform heavy image manipulations and to reduce the computational complexity of the model by doing the majority of processing at lower scales / resolutions. Nearly all models used standard image classification models in their encoder module extracting features from the input images. Teams \textit{Tencent GY-Lab}, \textit{SMART}, \textit{Airia-Team1} and \textit{CFL2} adopted MobileNets for this as they are already optimized for low-power devices and can achieve a very good runtime on the majority of mobile platforms. The best fidelity results were, however, obtained by team \textit{HIT-AIIA} that used the EfficientNet-B1 network for feature generation. To improve the models' accuracy, skip connections between the encoder and decoder blocks were added in almost all architectures. Another popular approach resulting in better depth prediction was to use knowledge distillation: a larger model was first trained for the same task, and then its outputs or intermediate features were used as additional targets for the final small network. In particular, this approach was used by the challenge winner, team \textit{Tencent GY-Lab}, that outperformed all other methods by a huge margin, being able to get both good fidelity scores and to achieve more than 10 FPS on the target Raspberry Pi 4 device. Notably, this solution is a magnitude faster than the FastDepth~\cite{wofk2019fastdepth} model known as one of the most efficient ones for this task.

To further benchmark the efficiency of the designed solutions, we additionally tested their performance on several popular smartphone chipsets. The runtime results demonstrated in Table~\ref{tab:runtime_results} were measured with the AI Benchmark using the TFLite GPU delegate~\cite{lee2019device} compatible with all mobile devices supporting OpenCL or OpenGL 3.0+.  In almost all cases, the runtime of the proposed networks is less than half a second except for the solution from \textit{3dv oppo}: due to the issues caused by PyTorch to TFLite conversion, it contains several ops supported neither by TFLite delegates nor by Android NNAPI, thus this model was executed on CPU, same as networks from \textit{Airia-Team1} and \textit{CFL2}. The solution from team \textit{Tencent GY-Lab} demonstrated more than 75 FPS on all considered SoCs, thus being able to generate depth maps in real-time on all modern chipsets, including the low-end ones. We can conclude that this architecture is now defining a new efficiency standard for depth estimation on mobile and embedded systems. The model from team \textit{HIT-AIIA}, demonstrating the best accuracy in this challenge, is able to achieve at least 7 FPS on all tested SoCs, thus being applicable for tasks where the precision of the predicted depth maps is critical. It should be also mentioned that all models were additionally tested on NPUs / DSPs of the considered chipsets, though the results were either the same or worse since not all TFLite layers and operations are currently optimized for specialized AI hardware.

\begin{figure*}[t!]
    \centering
    \includegraphics[width=0.66\textwidth]{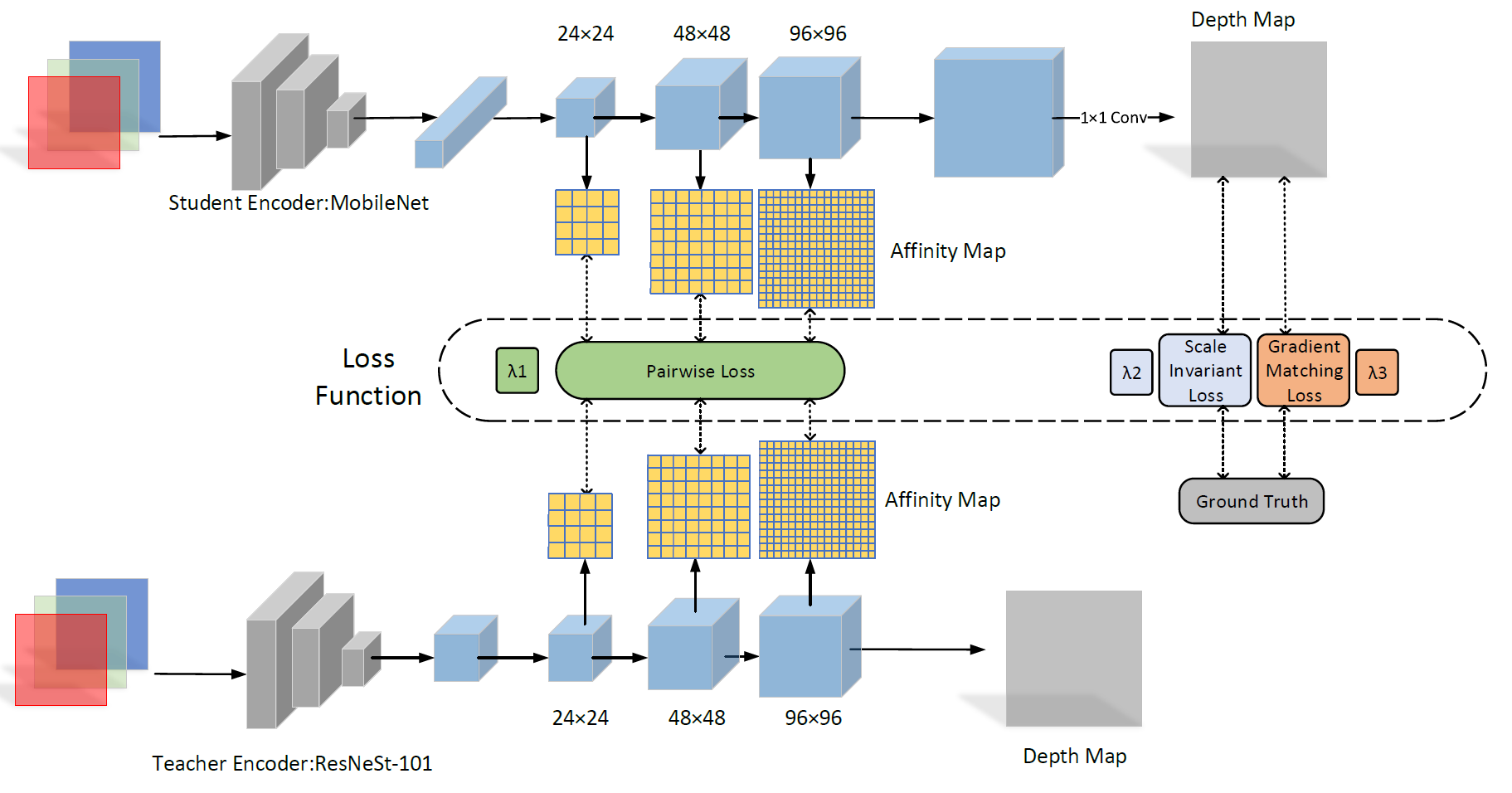}
    \caption{An overview of the knowledge distillation strategy used by team SMART.}
    \label{fig:kx_smart_training_strategy}
\end{figure*}

\section{Challenge Methods}
\label{sec:solutions}

\noindent This section describes solutions submitted by all teams participating in the final stage of the MAI 2021 Monocular Depth Estimation challenge.

\subsection{Tencent GY-Lab}

Team Tencent GY-Lab proposed a U-Net like architecture presented in Fig.~\ref{fig:light_architecture}, where a MobileNet-V3~\cite{howard2019searching} based encoder is used for dense feature extraction. To reduce the amount of computations, the input image is first resized from 640$\times$480 to 160$\times$128 pixels and then passed to the encoder module consisting of five blocks. The outputs of each block are processed by the Feature Fusion Module (FFM) that concatenates them with the decoder feature maps to get better fidelity results. The authors use one additional \textit{nearest neighbor} resizing layer on top of the model to upscale the output to the target resolution. Knowledge distillation~\cite{hinton2015distilling} is further used to improve the quality of the reconstructed depth maps: a bigger ViT-Large~\cite{dosovitskiy2020image} was first trained on the same dataset and then its features obtained before the last activation function were used to guide the smaller network. This process allowed to decrease the si-RMSE score from $0.3304$ to $0.3141$. The proposed model was therefore trained to minimize a combination of the distillation loss (computed as $L_2$ norm between its features from the last convolutional layer and the above mentioned features from the larger model), and the depth estimation loss proposed in~\cite{lee2019big}. The network parameters were optimized for 500 epochs using Adam~\cite{kingma2014adam} with a learning rate of $8e-3$ and a polynomial decay with a power of $0.9$. The model was implemented and trained with PyTorch and then converted to TensorFlow Lite using ONNX as an intermediate representation. A more detailed description of the proposed solution is provided in~\cite{zhang2021asimple}.

\subsection{SMART}

\begin{figure}[h!]
    \centering
    \includegraphics[width=1.0\linewidth]{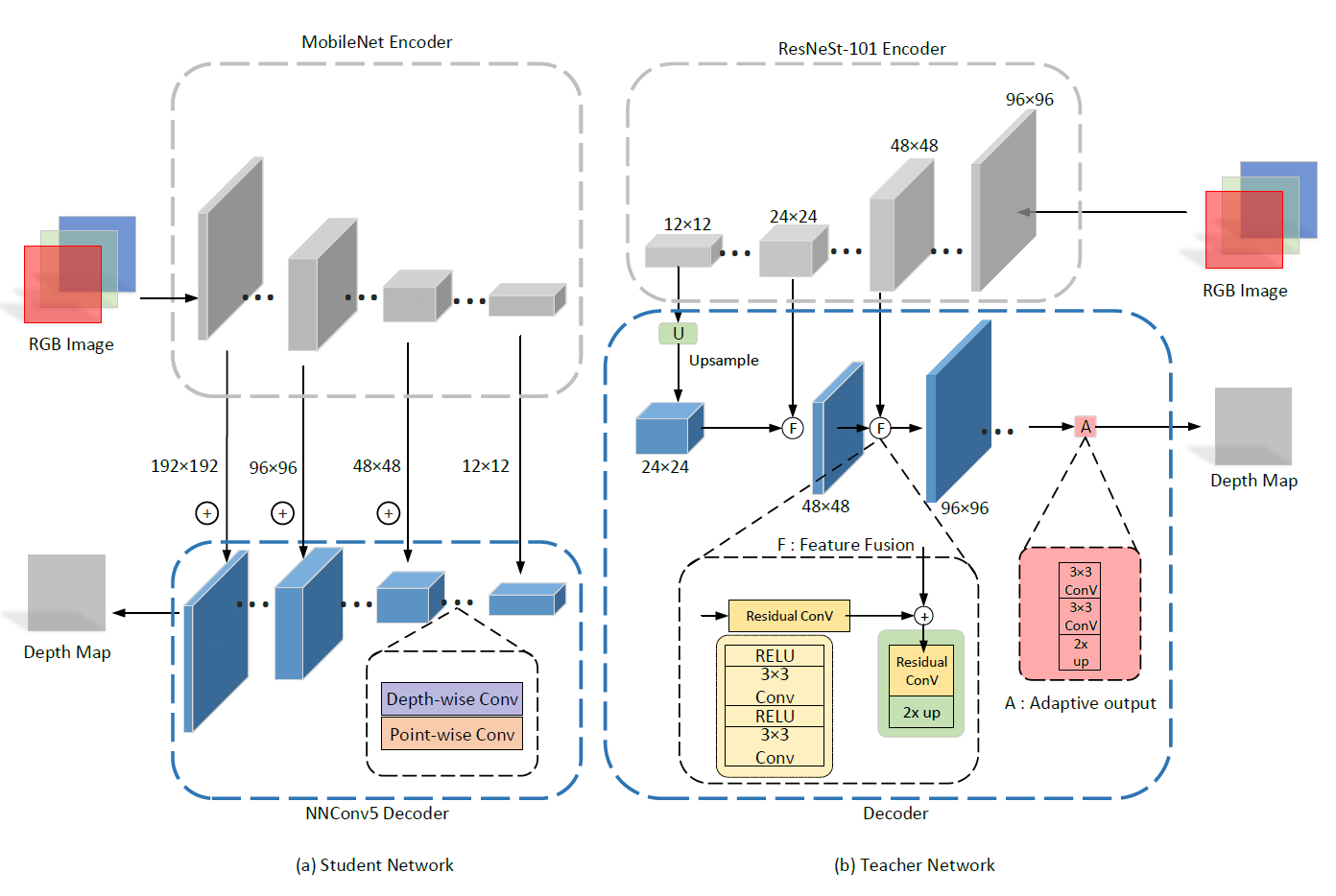}
    \caption{The architecture of the student and teacher models developed by team SMART.}
    \label{fig:kx_smart_architecture}
\end{figure}

Same as the previous solution, team SMART used a MobileNet-based encoder module for feature extraction and applied knowledge distillation to train the network. The architecture of the proposed solution is demonstrated in Fig.~\ref{fig:kx_smart_architecture}: the standard FastDepth~\cite{wofk2019fastdepth} architecture with a MobileNet-V1 backbone is used for the main (student) model. The larger teacher network consists of a ResNeSt-101~\cite{xie2017aggregated} based encoder and a decoder block~\cite{xian2018monocular} with the adaptive output layer on top of it. The representation ability of a pre-trained teacher model is transferred to the student network via knowledge distillation: a pairwise distillation loss is adopted to force the student network to output feature maps that are similar to the outputs of the corresponding layers of the teacher network. The distillation loss is computed in two steps (Fig.~\ref{fig:kx_smart_training_strategy}): let $F_t \in \mathbb{R}^{h\times w \times c_1}$ and $F_s \in \mathbb{R}^{h\times w \times c_2}$ be the feature maps with the same spatial resolution from the teacher and the student models, respectively, then the affinity maps are first computed as:
\begin{equation*}
    a_{ij}=\frac{f_i^T f_j}{(\lVert f_i \rVert_2 \times \lVert f_j \rVert_2)},
\end{equation*}
where $f$ denotes one row of the feature map ($F_t$ or $F_s$). Next, the mean square error is computed between the affinity maps obtained for student and teacher models:
\begin{equation*}
    \label{eq:distillation-loss}
    \mathcal{L}_{pa}(S,T)=\frac{1}{w\times h}\sum_i \sum_j (a_{ij}^s-a_{ij}^t)^2.
\end{equation*}

Besides that above knowledge distillation loss, two other metrics are used to train the student model. The scale invariant loss~\cite{eigen2014depth} is used to measure the discrepancy between the output of the student network and the ground truth depth map:
\begin{equation*}
    \mathcal{L}_s\left( d, d^* \right) =\frac{1}{n}\sum_i{g_{i}^{2}}-\frac{1}{n^2}(\sum_i{g_i})^2,
\end{equation*}
where $d$ and $d^*$ are the predicted and the ground truth depth maps, and $g_i=\log{d}_i-\log d_i^*$ is the corresponding error in log space. Finally, the scale-invariant gradient matching loss~\cite{ranftl2019towards} is defined as:
\begin{equation*}
    \mathcal{L}_{reg}\left( d, d^* \right)=\frac{1}{M}\sum_{k=1}^K\sum_{i=1}^M(|\nabla_x R_i^k|+|\nabla_y R_i^k|),
\end{equation*}
where $R_i = d - d^*$, and $R^k$ denotes the difference between the disparity maps at scale $k=1,2,3,4$ (the resolution of the feature maps is halved at each level). The final loss function is then defined as:
\begin{equation*}
    \mathcal{L} = 10 \cdot \mathcal{L}_s\left( d, d^* \right) + 0.1 \cdot \mathcal{L}_{reg}(d, d^*) + 1000 \cdot \mathcal{L}_{pa}(S,T).
\end{equation*}

The model was trained using Adam for 100 epochs with an initial learning rate of $1e-3$ and a polynomial decay with a power of $0.9$. A more detailed description of the model, design choices and training procedure is provided in~\cite{wang2021knowledge}.

\subsection{Airia-Team1}

\begin{figure}[h]
    \centering
    \includegraphics[width=1.0\linewidth]{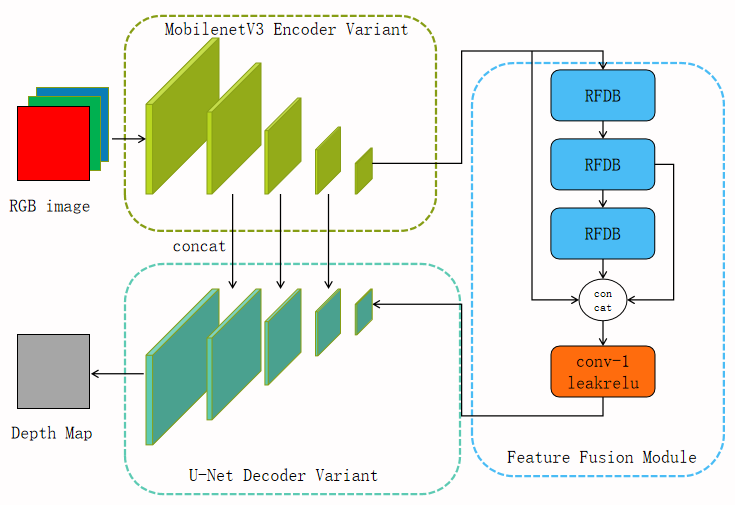}
    \includegraphics[width=0.7\linewidth]{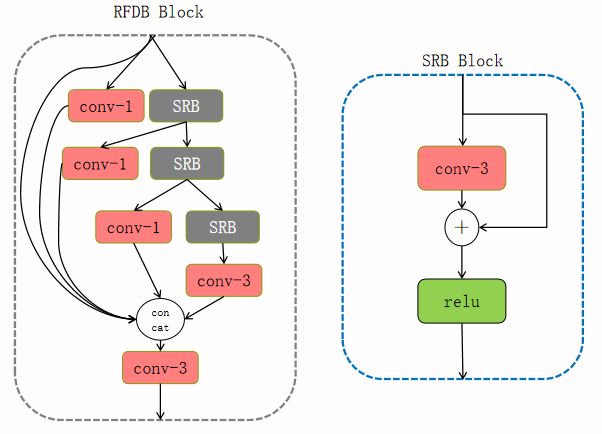}
    \caption{The architecture proposed by Airia-Team1 (top) and the structure of the RFDB block (bottom). \textit{Conv-1} and \textit{Conv-3} stands for 1$\times$1 and 3$\times$3 convolution, respectively.}
    \label{fig:airia_plot}
\end{figure}

Figure~\ref{fig:airia_plot} demonstrates the architecture developed by Airia-Team1. The authors proposed an encoder-decoder model, where MobileNet-V3~\cite{howard2019searching} network is used for feature extraction, same as in the previous two solutions. The resulting features are fed to three residual feature distillation blocks (RFDB), each one composed of three residual blocks (SRB) and several convolutional and concatenation layers. The refined features obtained after these blocks are then passed to a 5-layer decoder producing the final predictions, several skip connections are additionally used to speed-up the training. The pixel-wise depth loss~\cite{bhat2020adabins} was used as the target loss function. The model parameters were optimized using Adam with a learning rate of $1e-4$ multiplied by $0.6$ each $100$ epochs. A batch size of 8 was used during the training, random flips were additionally applied for data augmentation.

\subsection{YTL}

The authors proposed a U-net based architecture where the ResNet-18~\cite{he2016deep} model is used for feature extraction. The input RGB image was resized to 320$\times$240 resolution and then concatenated with an X/Y meshgrid (containing centered pixel coordinates) to form a 5-channel tensor passed to the model. The output of the model was also upsampled from 320$\times$240 to the target 640$\times$480 resolution using one \textit{bilinear resize} layer on top of it. The network was trained to minimize a combination of the Mean Absolute Error (MAE) and gradient losses using Adam optimizer.

\subsection{CFL2}

\begin{figure}[h]
    \centering
    \includegraphics[width=0.5\textwidth]{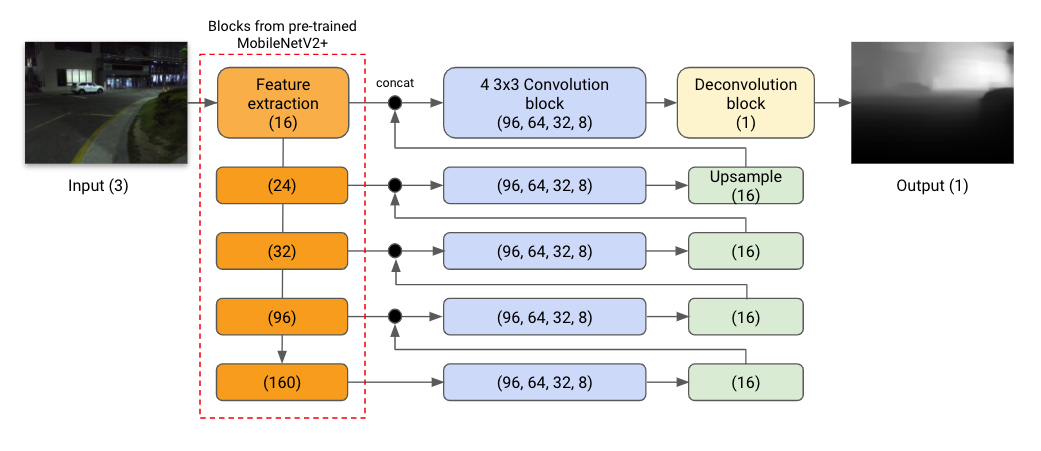}
    \caption{The PyDNet~\cite{aleotti2021real} architecture adopted by CFL2 team.}
    \label{fig:cfl_plot}
\end{figure}

Team CFL2 based its solution on the PyDNet~\cite{poggi2018towards,aleotti2021real} model. The input image was downscaled to 256$\times$256 pixels and then passed to the MobileNetV2~\cite{sandler2018mobilenetv2} encoder. While the original PyDNet model produces several outputs at multiple scales, the authors used only the highest one that corresponds to the target resolution to reduce the computational complexity of the model. Since the PyDNet is originally producing 128$\times$128px images, they were additionally upscaled to the target resolution using one \textit{bilinear resize} layer. The scale invariant data loss~\cite{eigen2014depth} and the scale-invariant gradient matching loss~\cite{ranftl2019towards} were used to train the model for 2M interations using Adam with a learning rate of $1e-4$.

\subsection{HIT-AIIA}

\begin{figure}[h]
    \centering
    \includegraphics[width=1.0\linewidth]{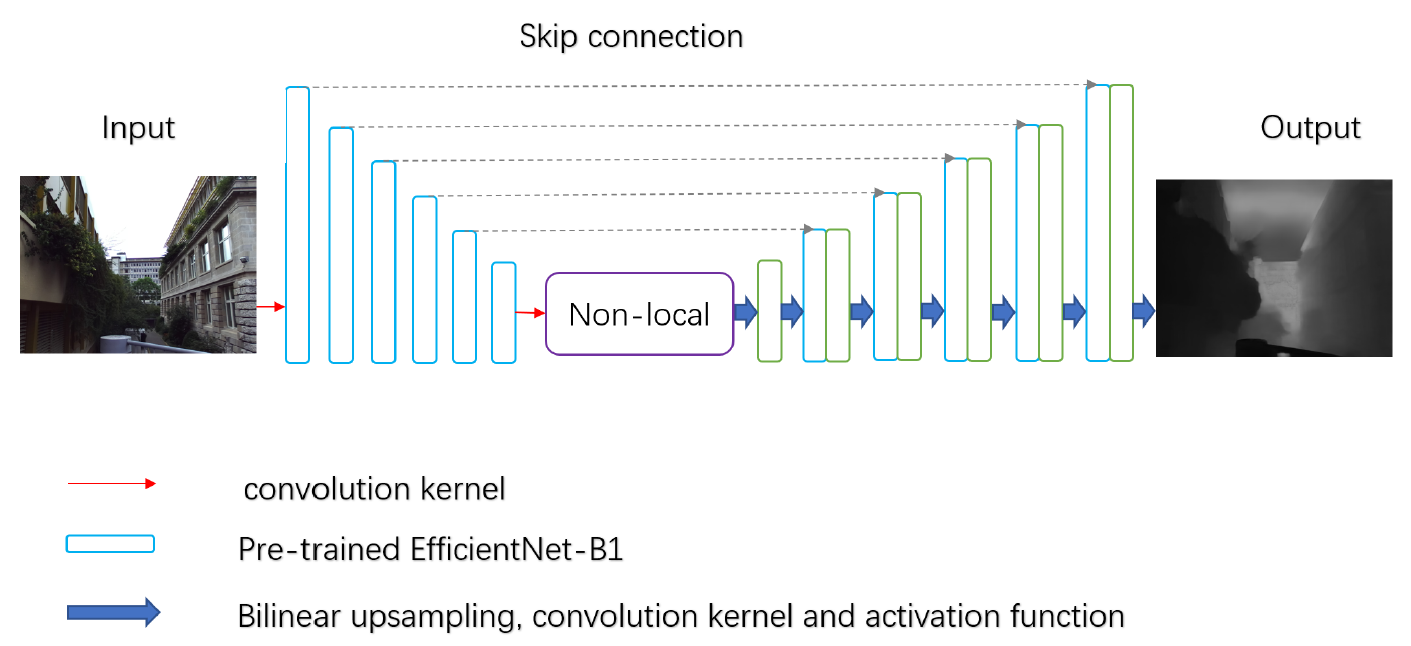}
    \caption{EfficientNet-based model proposed by team HIT-AIIA.}
    \label{fig:HITAIIA}
\end{figure}

The model proposed by HIT-AIIA is using the EfficientNet-B1 network~\cite{tan2019efficientnet} as an encoder to extract features from the input images (Fig.~\ref{fig:HITAIIA}). The outputs from its last layer are passed to the Non-Local block~\cite{wang2018non} that effectively improves the accuracy of the model. The authors used a combination of the bilinear upsampling, convolutional and \textit{Leaky ReLU} layers in the decoder module predicting the final depth map. Additional skip connections were added to speed-up the training process and improve the fidelity results. The model was trained to minimize RMSE loss function using Adam with a learning rate of $1e-4$ and a batch size of 6. Image mirroring and flipping as well as color alteration were used for data augmentation.

\subsection{weichi}

\begin{figure}[!h]
    \centering
    \includegraphics[width=0.5\textwidth]{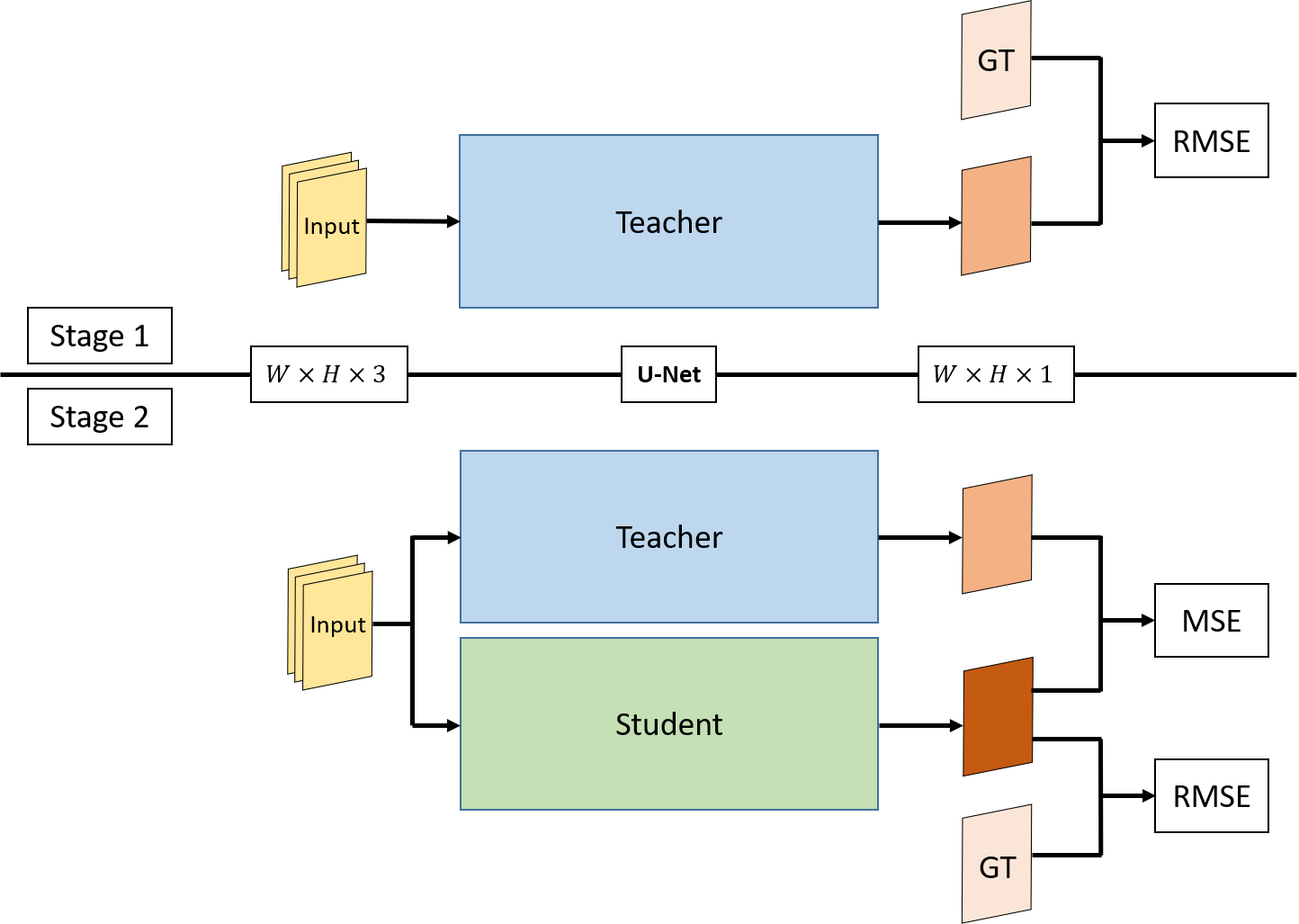}
    \caption{ An overview of the knowledge distillation strategy used by team weichi.}
    \label{fig:weichi_pipeline}
\end{figure}

Team weichi used the standard U-Net~\cite{ronneberger2015u} architecture with a reduced by a factor of 8 number of feature maps in each layer. Same as in~\cite{mangalam2018compressing}, the authors added batch normalization after each convolution in the encoder block. To improve the accuracy of the model, knowledge distillation~\cite{hinton2015distilling} was additionally applied during the training process: a larger U-Net model (with an increased number of channels) was first trained on the same dataset using the RMSE loss function. Next, the main student network was minimizing a combination of the RMSE loss between its outputs and the target depth maps, and the MSE loss between its outputs and the outputs of the larger network (Fig.~\ref{fig:weichi_pipeline}). Both models were trained using Adam optimizer with a learning rate of $5e-5$ and a batch size of 16.

\subsection{MonoVision Palace}

\begin{figure}[!h]
    \centering
    \includegraphics[width=0.5\textwidth]{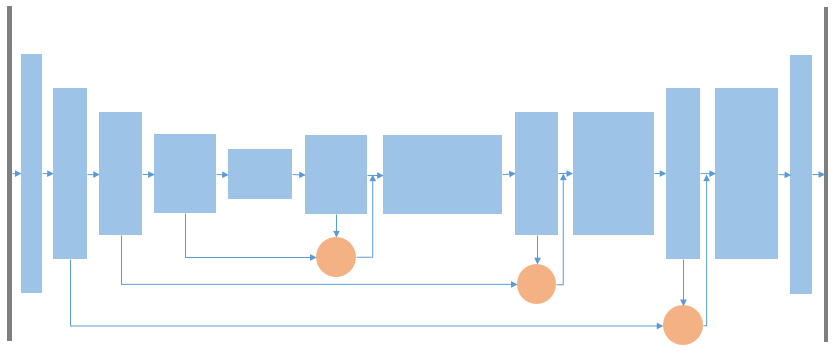}
    \caption{DA-UNet architecture proposed by MonoVision Palace.}
    \label{fig:mvp_architecture}
\end{figure}

Team MVP proposed a Depth Attention UNet (DA-UNet) architecture demonstrated in Fig.~\ref{fig:mvp_architecture}. The input image was first passed to the EfficientNet-Edge-TPU-S~\cite{gupta2019efficientnet} model with removed \textit{hard-swish} activations and \textit{squeeze-and-excitation} blocks to reduce the latency. Its outputs were then processed by the decoder block composed of convolution, upsampling, \textit{Leaky ReLU} and Gated Attention Blocks (GA)~\cite{oktay2018attention} where \textit{ReLU} and \textit{sigmoid} activations were replaced with \textit{Leaky ReLUs} and \textit{hard-sigmoid} ops, respectively. The model was trained using the same metrics as in~\cite{alhashim2018high}: the point-wise $L_1$ loss, the gradient $L_1$ loss, and the SSIM loss function. Adam was used to optimize the model parameters for 30 epochs with an initial learning rate of $1e-4$ reduced by a magnitude after the 20th and the 25th epoch.

\subsection{3dv oppo}

\begin{figure}[!h]
    \centering
    \includegraphics[width=1.0\linewidth]{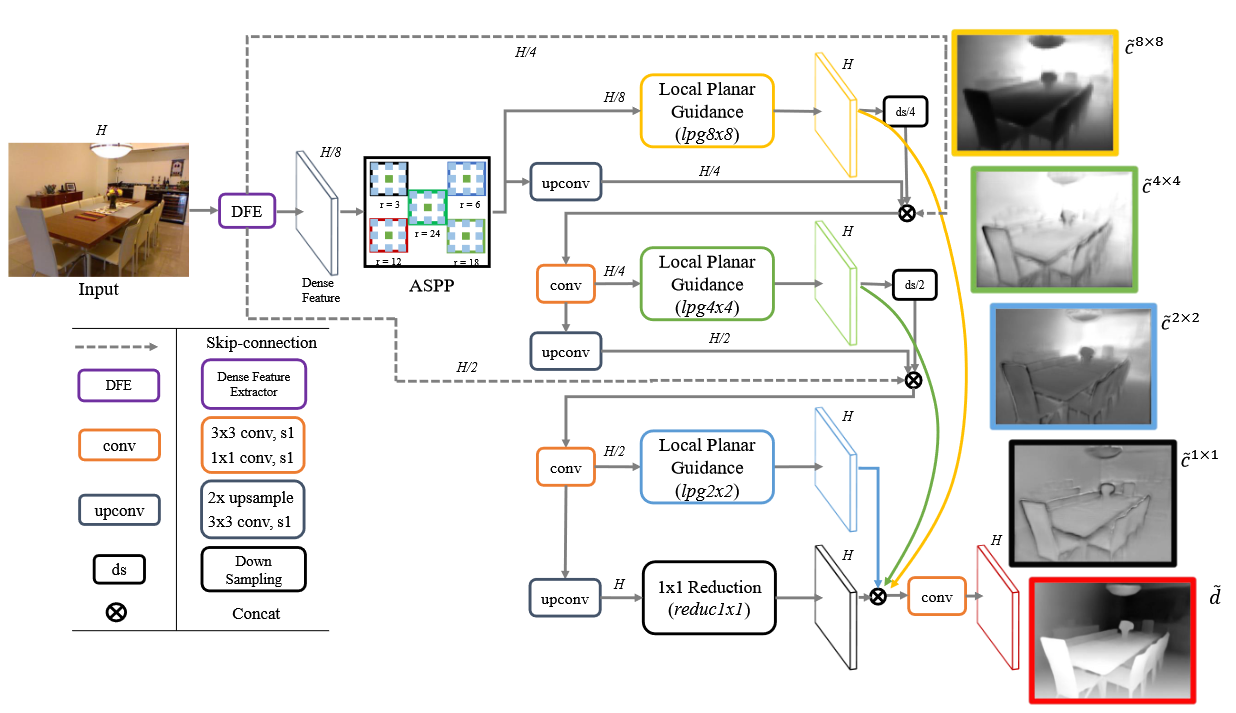}
    \caption{BTS network architecture adopted by 3dv oppo team.}
    \label{fig:3dvoppo}
\end{figure}

The authors directly used the BTS model~\cite{lee2019big} demonstrated in Fig~\ref{fig:3dvoppo}. This network is composed of the dense feature extractor (the ResNet model), the contextual information extractor (ASPP), the local planar guidance layers and their dense connection for final depth estimation. The same training setup and the target loss functions as in~\cite{lee2019big} was used except for the learning rate that was set to $5e-5$.

\subsection{MegaUe}
\begin{figure}[!h]
    \centering
    \includegraphics[width=0.94\linewidth]{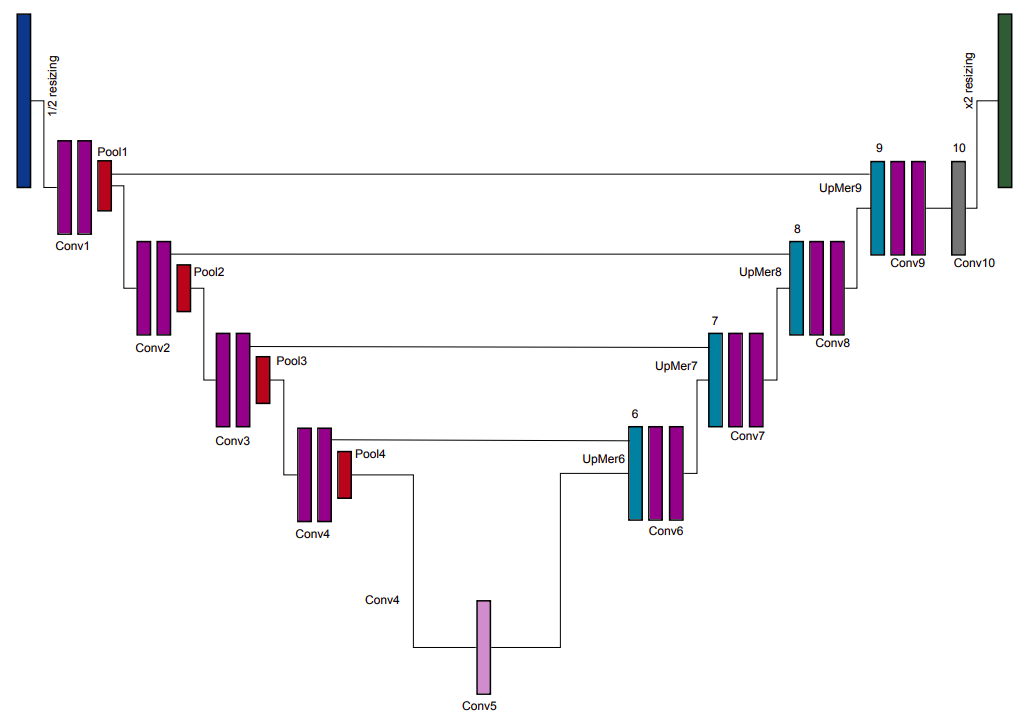}
    \caption{U-Net model proposed by MegaUe team.}
    \label{fig:MegaUe_architecture}
\end{figure}

Team MegaUe trained a standard U-Net like architecture (Fig.~\ref{fig:MegaUe_architecture}) with one additional 2x image downsampling and upsampling layers at the begging and at the top of the model, respectively. The model was first pre-trained on the MegaDepth dataset~\cite{li2018megadepth} using the same metrics as in the original paper: the ordinal, data and gradient matching losses. Then, the model was fine-tuned on the challenge data using the last two loss functions.

\section{Additional Literature}

An overview of the past challenges on mobile-related tasks together with the proposed solutions can be found in the following papers:

\begin{itemize}
\item Learned End-to-End ISP:\, \cite{ignatov2019aim,ignatov2020aim}
\item Perceptual Image Enhancement:\, \cite{ignatov2018pirm,ignatov2019ntire}
\item Image Super-Resolution:\, \cite{ignatov2018pirm,lugmayr2020ntire,cai2019ntire,timofte2018ntire}
\item Bokeh Effect Rendering:\, \cite{ignatov2019aimBokeh,ignatov2020aimBokeh}
\item Image Denoising:\, \cite{abdelhamed2020ntire,abdelhamed2019ntire}
\end{itemize}

\section*{Acknowledgements}

We thank Raspberry Pi (Trading) Ltd, AI Witchlabs and ETH Zurich (Computer Vision Lab), the organizers and sponsors of this Mobile AI 2021 challenge.

\appendix
\section{Teams and Affiliations}
\label{sec:apd:team}

\bigskip

\subsection*{Mobile AI 2021 Team}
\noindent\textit{\textbf{Title: }}\\ Mobile AI 2021 Challenge on Single-Image Depth Estimation on Mobile Devices\\
\noindent\textit{\textbf{Members:}}\\ Andrey Ignatov$^{1,3}$ \textit{(andrey@vision.ee.ethz.ch)}, Grigory Malivenko \textit{(grigory.malivenko@gmail.com)}, David Plowman$^2$ \textit{(david.plowman@raspberrypi.com)}, Samarth Shukla$^1$ \textit{(samarth.shukla@vision.ee.ethz.ch)}, Radu Timofte$^{1,3}$ \textit{(radu.timofte@vision.ee.ethz.ch)}\\
\noindent\textit{\textbf{Affiliations: }}\\
$^1$ Computer Vision Lab, ETH Zurich, Switzerland\\
$^2$ Raspberry Pi (Trading) Ltd\\
$^3$ AI Witchlabs, Switzerland\\

\subsection*{Tencent GY-Lab}
\noindent\textit{\textbf{Title:}}\\ A Simple Baseline for Fast and Accurate Depth Estimation on Mobile Devices~\cite{zhang2021asimple} \\
\noindent\textit{\textbf{Members: }}\\ \textit{Ziyu Zhang (parkzyzhang@tencent.com)}, Yicheng Wang, Zilong Huang, Guozhong Luo, Gang Yu, Bin Fu \\
\noindent\textit{\textbf{Affiliations: }}\\
Tencent GY-Lab, China\\

\subsection*{SMART}
\noindent\textit{\textbf{Title:}}\\ Knowledge Distillation for Fast and Accurate Monocular Depth Estimation on Mobile Devices~\cite{wang2021knowledge} \\
\noindent\textit{\textbf{Members: }}\\ \textit{Yiran Wang (wangyiran@hust.edu.cn)}, Xingyi Li, Min Shi, Ke Xian, Zhiguo Cao \\
\noindent\textit{\textbf{Affiliations: }}\\
Key Laboratory of Image Processing and Intelligent Control, Ministry of Education, School of Artificial Intelligence and Automation, Huazhong University of Science and Technology, China\\

\subsection*{Airia-Team1}
\noindent\textit{\textbf{Title:}}\\ Monocular Depth Estimation based on MobileNetV3Small \\
\noindent\textit{\textbf{Members: }}\\ \textit{ Jin-Hua Du (2982192572@qq.com)}, Pei-Lin Wu, Chao Ge  \\
\noindent\textit{\textbf{Affiliations: }}\\
Nanjing Artificial Intelligence Chip Research, Institute of Automation, Chinese Academy of Sciences, China\\

\subsection*{YTL}
\noindent\textit{\textbf{Title:}}\\ U-Net with Pixel Position Encoding for Monocular Depth Estimation \\
\noindent\textit{\textbf{Members: }}\\ \textit{ Jiaoyang Yao (jiaoyangyao@gmail.com)}, Fangwen Tu, Bo
Li  \\
\noindent\textit{\textbf{Affiliations: }}\\
Black Sesame Technologies Inc., Singapore \\

\subsection*{CFL2}
\noindent\textit{\textbf{Title:}}\\ Lightfast Depth Estimation \\
\noindent\textit{\textbf{Members: }}\\ \textit{ Jung Eun Yoo (jey920@kaist.ac.kr)}, Kwanggyoon Seo \\
\noindent\textit{\textbf{Affiliations: }}\\
Visual Media Lab, KAIST, South Korea\\

\subsection*{HIT-AIIA}
\noindent\textit{\textbf{Title:}}\\ EfficientNet Encoder with Non-Local Module for Monocular Depth Estimation \\
\noindent\textit{\textbf{Members: }}\\ \textit{ Jialei Xu (20S003044@stu.hit.edu.cn)}, Zhenyu Li, Xianming Liu, Junjun Jiang \\
\noindent\textit{\textbf{Affiliations: }}\\
Harbin Institute of Technology, China \\
Peng Cheng Laboratory, China \\

\subsection*{weichi}
\noindent\textit{\textbf{Title:}}\\ Distillation on UNet \\
\noindent\textit{\textbf{Members: }}\\ \textit{ Wei-Chi Chen (ne6094041@gs.ncku.edu.tw)} \\
\noindent\textit{\textbf{Affiliations: }}\\
Multimedia and Computer Vision Laboratory, National Cheng Kung University, Taiwan \\
\url{http://mmcv.csie.ncku.edu.tw/}\\

\subsection*{MVP - MonoVision Palace}
\noindent\textit{\textbf{Title:}}\\ DA-UNet: Depth Attention UNet for Monocular Depth Estimation \\
\noindent\textit{\textbf{Members: }}\\ \textit{ Shayan Joya (joya.shayan@gmail.com)} \\
\noindent\textit{\textbf{Affiliations: }}\\
Samsung Research UK, United Kingdom \\

\subsection*{3dv oppo}
\noindent\textit{\textbf{Title:}}\\ Accurate Monocular Depth Estimation Using BTS \\
\noindent\textit{\textbf{Members: }}\\ \textit{ Huanhuan Fan (fanhuanhuan@oppo.com)}, Zhaobing Kang, Ang Li, Tianpeng Feng, Yang Liu, Chuannan Sheng, Jian Yin \\
\noindent\textit{\textbf{Affiliations: }}\\
OPPO Research Institute, China \\

\subsection*{MegaUe}
\noindent\textit{\textbf{Title:}}\\ Mega-Udepth for Monocular Depth Estimation \\
\noindent\textit{\textbf{Members: }}\\ \textit{ Fausto T. Benavides (fausto.tapiabenavides@gmail.com)} \\
\noindent\textit{\textbf{Affiliations: }}\\
ETH Zurich, Switzerland \\

{\small
\bibliographystyle{ieee_fullname}

\begin{thebibliography}{10}\itemsep=-1pt

\bibitem{abdelhamed2020ntire}
Abdelrahman Abdelhamed, Mahmoud Afifi, Radu Timofte, and Michael~S Brown.
\newblock Ntire 2020 challenge on real image denoising: Dataset, methods and
  results.
\newblock In {\em Proceedings of the IEEE/CVF Conference on Computer Vision and
  Pattern Recognition Workshops}, pages 496--497, 2020.

\bibitem{abdelhamed2019ntire}
Abdelrahman Abdelhamed, Radu Timofte, and Michael~S Brown.
\newblock Ntire 2019 challenge on real image denoising: Methods and results.
\newblock In {\em Proceedings of the IEEE/CVF Conference on Computer Vision and
  Pattern Recognition Workshops}, pages 0--0, 2019.

\bibitem{aleotti2021real}
Filippo Aleotti, Giulio Zaccaroni, Luca Bartolomei, Matteo Poggi, Fabio Tosi,
  and Stefano Mattoccia.
\newblock Real-time single image depth perception in the wild with handheld
  devices.
\newblock {\em Sensors}, 21(1):15, 2021.

\bibitem{alhashim2018high}
Ibraheem Alhashim and Peter Wonka.
\newblock High quality monocular depth estimation via transfer learning.
\newblock {\em arXiv preprint arXiv:1812.11941}, 2018.

\bibitem{NNAPI2021}
Android Neural~Networks API.
\newblock \url{https://developer.android.com/ndk/guides/neuralnetworks}.

\bibitem{AIBenchmark202104}
AI~Benchmark Archive.
\newblock
  \url{http://web.archive.org/web/20210425131428/https://ai-benchmark.com/ranking_processors.html}.

\bibitem{bhat2020adabins}
Shariq~Farooq Bhat, Ibraheem Alhashim, and Peter Wonka.
\newblock Adabins: Depth estimation using adaptive bins.
\newblock {\em arXiv preprint arXiv:2011.14141}, 2020.

\bibitem{cai2019ntire}
Jianrui Cai, Shuhang Gu, Radu Timofte, and Lei Zhang.
\newblock Ntire 2019 challenge on real image super-resolution: Methods and
  results.
\newblock In {\em Proceedings of the IEEE/CVF Conference on Computer Vision and
  Pattern Recognition Workshops}, pages 0--0, 2019.

\bibitem{cai2020zeroq}
Yaohui Cai, Zhewei Yao, Zhen Dong, Amir Gholami, Michael~W Mahoney, and Kurt
  Keutzer.
\newblock Zeroq: A novel zero shot quantization framework.
\newblock In {\em Proceedings of the IEEE/CVF Conference on Computer Vision and
  Pattern Recognition}, pages 13169--13178, 2020.

\bibitem{chen2016single}
Weifeng Chen, Zhao Fu, Dawei Yang, and Jia Deng.
\newblock Single-image depth perception in the wild.
\newblock {\em arXiv preprint arXiv:1604.03901}, 2016.

\bibitem{chiang2020deploying}
Cheng-Ming Chiang, Yu Tseng, Yu-Syuan Xu, Hsien-Kai Kuo, Yi-Min Tsai, Guan-Yu
  Chen, Koan-Sin Tan, Wei-Ting Wang, Yu-Chieh Lin, Shou-Yao~Roy Tseng, et~al.
\newblock Deploying image deblurring across mobile devices: A perspective of
  quality and latency.
\newblock In {\em Proceedings of the IEEE/CVF Conference on Computer Vision and
  Pattern Recognition Workshops}, pages 502--503, 2020.

\bibitem{TFLiteDelegates2021}
TensorFlow~Lite delegates.
\newblock \url{https://www.tensorflow.org/lite/performance/delegates}.

\bibitem{dosovitskiy2020image}
Alexey Dosovitskiy, Lucas Beyer, Alexander Kolesnikov, Dirk Weissenborn,
  Xiaohua Zhai, Thomas Unterthiner, Mostafa Dehghani, Matthias Minderer, Georg
  Heigold, Sylvain Gelly, et~al.
\newblock An image is worth 16x16 words: Transformers for image recognition at
  scale.
\newblock {\em arXiv preprint arXiv:2010.11929}, 2020.

\bibitem{eigen2014depth}
David Eigen, Christian Puhrsch, and Rob Fergus.
\newblock Depth map prediction from a single image using a multi-scale deep
  network.
\newblock {\em arXiv preprint arXiv:1406.2283}, 2014.

\bibitem{garg2016unsupervised}
Ravi Garg, Vijay~Kumar Bg, Gustavo Carneiro, and Ian Reid.
\newblock Unsupervised cnn for single view depth estimation: Geometry to the
  rescue.
\newblock In {\em European conference on computer vision}, pages 740--756.
  Springer, 2016.

\bibitem{godard2019digging}
Cl{\'e}ment Godard, Oisin Mac~Aodha, Michael Firman, and Gabriel~J Brostow.
\newblock Digging into self-supervised monocular depth estimation.
\newblock In {\em Proceedings of the IEEE/CVF International Conference on
  Computer Vision}, pages 3828--3838, 2019.

\bibitem{gupta2019efficientnet}
Suyog Gupta and Mingxing Tan.
\newblock Efficientnet-edgetpu: Creating accelerator-optimized neural networks
  with automl, 2019.

\bibitem{he2016deep}
Kaiming He, Xiangyu Zhang, Shaoqing Ren, and Jian Sun.
\newblock Deep residual learning for image recognition.
\newblock In {\em Proceedings of the IEEE conference on computer vision and
  pattern recognition}, pages 770--778, 2016.

\bibitem{hinton2015distilling}
Geoffrey Hinton, Oriol Vinyals, and Jeff Dean.
\newblock Distilling the knowledge in a neural network.
\newblock {\em arXiv preprint arXiv:1503.02531}, 2015.

\bibitem{howard2019searching}
Andrew Howard, Mark Sandler, Grace Chu, Liang-Chieh Chen, Bo Chen, Mingxing
  Tan, Weijun Wang, Yukun Zhu, Ruoming Pang, Vijay Vasudevan, et~al.
\newblock Searching for mobilenetv3.
\newblock In {\em Proceedings of the IEEE/CVF International Conference on
  Computer Vision}, pages 1314--1324, 2019.

\bibitem{ignatov2021fastDenoising}
Andrey Ignatov, Kim Byeoung-su, and Radu Timofte.
\newblock Fast camera image denoising on mobile gpus with deep learning, mobile
  ai 2021 challenge: Report.
\newblock In {\em Proceedings of the IEEE/CVF Conference on Computer Vision and
  Pattern Recognition Workshops}, pages 0--0, 2021.

\bibitem{ignatov2021learned}
Andrey Ignatov, Jimmy Chiang, Hsien-Kai Kuo, Anastasia Sycheva, and Radu
  Timofte.
\newblock Learned smartphone isp on mobile npus with deep learning, mobile ai
  2021 challenge: Report.
\newblock In {\em Proceedings of the IEEE/CVF Conference on Computer Vision and
  Pattern Recognition Workshops}, pages 0--0, 2021.

\bibitem{ignatov2017dslr}
Andrey Ignatov, Nikolay Kobyshev, Radu Timofte, Kenneth Vanhoey, and Luc
  Van~Gool.
\newblock Dslr-quality photos on mobile devices with deep convolutional
  networks.
\newblock In {\em Proceedings of the IEEE International Conference on Computer
  Vision}, pages 3277--3285, 2017.

\bibitem{ignatov2018wespe}
Andrey Ignatov, Nikolay Kobyshev, Radu Timofte, Kenneth Vanhoey, and Luc
  Van~Gool.
\newblock Wespe: weakly supervised photo enhancer for digital cameras.
\newblock In {\em Proceedings of the IEEE Conference on Computer Vision and
  Pattern Recognition Workshops}, pages 691--700, 2018.

\bibitem{ignatov2021fastSceneDetection}
Andrey Ignatov, Grigory Malivenko, and Radu Timofte.
\newblock Fast and accurate quantized camera scene detection on smartphones,
  mobile ai 2021 challenge: Report.
\newblock In {\em Proceedings of the IEEE/CVF Conference on Computer Vision and
  Pattern Recognition Workshops}, pages 0--0, 2021.

\bibitem{ignatov2020rendering}
Andrey Ignatov, Jagruti Patel, and Radu Timofte.
\newblock Rendering natural camera bokeh effect with deep learning.
\newblock In {\em Proceedings of the IEEE/CVF Conference on Computer Vision and
  Pattern Recognition Workshops}, pages 418--419, 2020.

\bibitem{ignatov2019aimBokeh}
Andrey Ignatov, Jagruti Patel, Radu Timofte, Bolun Zheng, Xin Ye, Li Huang,
  Xiang Tian, Saikat Dutta, Kuldeep Purohit, Praveen Kandula, et~al.
\newblock Aim 2019 challenge on bokeh effect synthesis: Methods and results.
\newblock In {\em 2019 IEEE/CVF International Conference on Computer Vision
  Workshop (ICCVW)}, pages 3591--3598. IEEE, 2019.

\bibitem{romero2021real}
Andrey Ignatov, Andres Romero, Heewon Kim, and Radu Timofte.
\newblock Real-time video super-resolution on smartphones with deep learning,
  mobile ai 2021 challenge: Report.
\newblock In {\em Proceedings of the IEEE/CVF Conference on Computer Vision and
  Pattern Recognition Workshops}, pages 0--0, 2021.

\bibitem{ignatov2019ntire}
Andrey Ignatov and Radu Timofte.
\newblock Ntire 2019 challenge on image enhancement: Methods and results.
\newblock In {\em Proceedings of the IEEE/CVF Conference on Computer Vision and
  Pattern Recognition Workshops}, pages 0--0, 2019.

\bibitem{ignatov2018ai}
Andrey Ignatov, Radu Timofte, William Chou, Ke Wang, Max Wu, Tim Hartley, and
  Luc Van~Gool.
\newblock Ai benchmark: Running deep neural networks on android smartphones.
\newblock In {\em Proceedings of the European Conference on Computer Vision
  (ECCV) Workshops}, pages 0--0, 2018.

\bibitem{ignatov2021real}
Andrey Ignatov, Radu Timofte, Maurizio Denna, and Abdel Younes.
\newblock Real-time quantized image super-resolution on mobile npus, mobile ai
  2021 challenge: Report.
\newblock In {\em Proceedings of the IEEE/CVF Conference on Computer Vision and
  Pattern Recognition Workshops}, pages 0--0, 2021.

\bibitem{ignatov2019aim}
Andrey Ignatov, Radu Timofte, Sung-Jea Ko, Seung-Wook Kim, Kwang-Hyun Uhm,
  Seo-Won Ji, Sung-Jin Cho, Jun-Pyo Hong, Kangfu Mei, Juncheng Li, et~al.
\newblock Aim 2019 challenge on raw to rgb mapping: Methods and results.
\newblock In {\em 2019 IEEE/CVF International Conference on Computer Vision
  Workshop (ICCVW)}, pages 3584--3590. IEEE, 2019.

\bibitem{ignatov2019ai}
Andrey Ignatov, Radu Timofte, Andrei Kulik, Seungsoo Yang, Ke Wang, Felix Baum,
  Max Wu, Lirong Xu, and Luc Van~Gool.
\newblock Ai benchmark: All about deep learning on smartphones in 2019.
\newblock In {\em 2019 IEEE/CVF International Conference on Computer Vision
  Workshop (ICCVW)}, pages 3617--3635. IEEE, 2019.

\bibitem{ignatov2020aimBokeh}
Andrey Ignatov, Radu Timofte, Ming Qian, Congyu Qiao, Jiamin Lin, Zhenyu Guo,
  Chenghua Li, Cong Leng, Jian Cheng, Juewen Peng, et~al.
\newblock Aim 2020 challenge on rendering realistic bokeh.
\newblock In {\em European Conference on Computer Vision}, pages 213--228.
  Springer, 2020.

\bibitem{ignatov2018pirm}
Andrey Ignatov, Radu Timofte, Thang Van~Vu, Tung Minh~Luu, Trung X~Pham, Cao
  Van~Nguyen, Yongwoo Kim, Jae-Seok Choi, Munchurl Kim, Jie Huang, et~al.
\newblock Pirm challenge on perceptual image enhancement on smartphones:
  Report.
\newblock In {\em Proceedings of the European Conference on Computer Vision
  (ECCV) Workshops}, pages 0--0, 2018.

\bibitem{ignatov2020aim}
Andrey Ignatov, Radu Timofte, Zhilu Zhang, Ming Liu, Haolin Wang, Wangmeng Zuo,
  Jiawei Zhang, Ruimao Zhang, Zhanglin Peng, Sijie Ren, et~al.
\newblock Aim 2020 challenge on learned image signal processing pipeline.
\newblock {\em arXiv preprint arXiv:2011.04994}, 2020.

\bibitem{ignatov2020replacing}
Andrey Ignatov, Luc Van~Gool, and Radu Timofte.
\newblock Replacing mobile camera isp with a single deep learning model.
\newblock In {\em Proceedings of the IEEE/CVF Conference on Computer Vision and
  Pattern Recognition Workshops}, pages 536--537, 2020.

\bibitem{ignatov2020controlling}
Dmitry Ignatov and Andrey Ignatov.
\newblock Controlling information capacity of binary neural network.
\newblock {\em Pattern Recognition Letters}, 138:276--281, 2020.

\bibitem{jacob2018quantization}
Benoit Jacob, Skirmantas Kligys, Bo Chen, Menglong Zhu, Matthew Tang, Andrew
  Howard, Hartwig Adam, and Dmitry Kalenichenko.
\newblock Quantization and training of neural networks for efficient
  integer-arithmetic-only inference.
\newblock In {\em Proceedings of the IEEE Conference on Computer Vision and
  Pattern Recognition}, pages 2704--2713, 2018.

\bibitem{jain2019trained}
Sambhav~R Jain, Albert Gural, Michael Wu, and Chris~H Dick.
\newblock Trained quantization thresholds for accurate and efficient
  fixed-point inference of deep neural networks.
\newblock {\em arXiv preprint arXiv:1903.08066}, 2019.

\bibitem{kingma2014adam}
Diederik~P Kingma and Jimmy Ba.
\newblock Adam: A method for stochastic optimization.
\newblock {\em arXiv preprint arXiv:1412.6980}, 2014.

\bibitem{laina2016deeper}
Iro Laina, Christian Rupprecht, Vasileios Belagiannis, Federico Tombari, and
  Nassir Navab.
\newblock Deeper depth prediction with fully convolutional residual networks.
\newblock In {\em 2016 Fourth international conference on 3D vision (3DV)},
  pages 239--248. IEEE, 2016.

\bibitem{lee2019device}
Juhyun Lee, Nikolay Chirkov, Ekaterina Ignasheva, Yury Pisarchyk, Mogan Shieh,
  Fabio Riccardi, Raman Sarokin, Andrei Kulik, and Matthias Grundmann.
\newblock On-device neural net inference with mobile gpus.
\newblock {\em arXiv preprint arXiv:1907.01989}, 2019.

\bibitem{lee2019big}
Jin~Han Lee, Myung-Kyu Han, Dong~Wook Ko, and Il~Hong Suh.
\newblock From big to small: Multi-scale local planar guidance for monocular
  depth estimation.
\newblock {\em arXiv preprint arXiv:1907.10326}, 2019.

\bibitem{li2019learning}
Yawei Li, Shuhang Gu, Luc~Van Gool, and Radu Timofte.
\newblock Learning filter basis for convolutional neural network compression.
\newblock In {\em Proceedings of the IEEE/CVF International Conference on
  Computer Vision}, pages 5623--5632, 2019.

\bibitem{li2018megadepth}
Zhengqi Li and Noah Snavely.
\newblock Megadepth: Learning single-view depth prediction from internet
  photos.
\newblock In {\em Proceedings of the IEEE Conference on Computer Vision and
  Pattern Recognition}, pages 2041--2050, 2018.

\bibitem{liu2015deep}
Fayao Liu, Chunhua Shen, and Guosheng Lin.
\newblock Deep convolutional neural fields for depth estimation from a single
  image.
\newblock In {\em Proceedings of the IEEE conference on computer vision and
  pattern recognition}, pages 5162--5170, 2015.

\bibitem{liu2015learning}
Fayao Liu, Chunhua Shen, Guosheng Lin, and Ian Reid.
\newblock Learning depth from single monocular images using deep convolutional
  neural fields.
\newblock {\em IEEE transactions on pattern analysis and machine intelligence},
  38(10):2024--2039, 2015.

\bibitem{liu2019metapruning}
Zechun Liu, Haoyuan Mu, Xiangyu Zhang, Zichao Guo, Xin Yang, Kwang-Ting Cheng,
  and Jian Sun.
\newblock Metapruning: Meta learning for automatic neural network channel
  pruning.
\newblock In {\em Proceedings of the IEEE/CVF International Conference on
  Computer Vision}, pages 3296--3305, 2019.

\bibitem{liu2018bi}
Zechun Liu, Baoyuan Wu, Wenhan Luo, Xin Yang, Wei Liu, and Kwang-Ting Cheng.
\newblock Bi-real net: Enhancing the performance of 1-bit cnns with improved
  representational capability and advanced training algorithm.
\newblock In {\em Proceedings of the European conference on computer vision
  (ECCV)}, pages 722--737, 2018.

\bibitem{lugmayr2020ntire}
Andreas Lugmayr, Martin Danelljan, and Radu Timofte.
\newblock Ntire 2020 challenge on real-world image super-resolution: Methods
  and results.
\newblock In {\em Proceedings of the IEEE/CVF Conference on Computer Vision and
  Pattern Recognition Workshops}, pages 494--495, 2020.

\bibitem{mangalam2018compressing}
Karttikeya Mangalam and Mathieu Salzamann.
\newblock On compressing u-net using knowledge distillation.
\newblock {\em arXiv preprint arXiv:1812.00249}, 2018.

\bibitem{obukhov2020t}
Anton Obukhov, Maxim Rakhuba, Stamatios Georgoulis, Menelaos Kanakis, Dengxin
  Dai, and Luc Van~Gool.
\newblock T-basis: a compact representation for neural networks.
\newblock In {\em International Conference on Machine Learning}, pages
  7392--7404. PMLR, 2020.

\bibitem{oktay2018attention}
Ozan Oktay, Jo Schlemper, Loic~Le Folgoc, Matthew Lee, Mattias Heinrich,
  Kazunari Misawa, Kensaku Mori, Steven McDonagh, Nils~Y Hammerla, Bernhard
  Kainz, et~al.
\newblock Attention u-net: Learning where to look for the pancreas.
\newblock {\em arXiv preprint arXiv:1804.03999}, 2018.

\bibitem{ortiz2018depth}
Luis~Enrique Ortiz, Elizabeth~V Cabrera, and Luiz~M Gon{\c{c}}alves.
\newblock Depth data error modeling of the zed 3d vision sensor from
  stereolabs.
\newblock {\em ELCVIA: electronic letters on computer vision and image
  analysis}, 17(1):0001--15, 2018.

\bibitem{poggi2018towards}
Matteo Poggi, Filippo Aleotti, Fabio Tosi, and Stefano Mattoccia.
\newblock Towards real-time unsupervised monocular depth estimation on cpu.
\newblock In {\em 2018 IEEE/RSJ International Conference on Intelligent Robots
  and Systems (IROS)}, pages 5848--5854. IEEE, 2018.

\bibitem{ranftl2019towards}
Ren{\'e} Ranftl, Katrin Lasinger, David Hafner, Konrad Schindler, and Vladlen
  Koltun.
\newblock Towards robust monocular depth estimation: Mixing datasets for
  zero-shot cross-dataset transfer.
\newblock {\em arXiv preprint arXiv:1907.01341}, 2019.

\bibitem{ronneberger2015u}
Olaf Ronneberger, Philipp Fischer, and Thomas Brox.
\newblock U-net: Convolutional networks for biomedical image segmentation.
\newblock In {\em International Conference on Medical image computing and
  computer-assisted intervention}, pages 234--241. Springer, 2015.

\bibitem{sandler2018mobilenetv2}
Mark Sandler, Andrew Howard, Menglong Zhu, Andrey Zhmoginov, and Liang-Chieh
  Chen.
\newblock Mobilenetv2: Inverted residuals and linear bottlenecks.
\newblock In {\em Proceedings of the IEEE conference on computer vision and
  pattern recognition}, pages 4510--4520, 2018.

\bibitem{tan2019mnasnet}
Mingxing Tan, Bo Chen, Ruoming Pang, Vijay Vasudevan, Mark Sandler, Andrew
  Howard, and Quoc~V Le.
\newblock Mnasnet: Platform-aware neural architecture search for mobile.
\newblock In {\em Proceedings of the IEEE/CVF Conference on Computer Vision and
  Pattern Recognition}, pages 2820--2828, 2019.

\bibitem{tan2019efficientnet}
Mingxing Tan and Quoc Le.
\newblock Efficientnet: Rethinking model scaling for convolutional neural
  networks.
\newblock In {\em International Conference on Machine Learning}, pages
  6105--6114. PMLR, 2019.

\bibitem{TensorFlowLite2021}
TensorFlow-Lite.
\newblock \url{https://www.tensorflow.org/lite}.

\bibitem{TensorFlowLite2021Linux}
TensorFlow-Lite.
\newblock \url{https://www.tensorflow.org/lite/guide/python}.

\bibitem{timofte2018ntire}
Radu Timofte, Shuhang Gu, Jiqing Wu, and Luc Van~Gool.
\newblock Ntire 2018 challenge on single image super-resolution: Methods and
  results.
\newblock In {\em Proceedings of the IEEE conference on computer vision and
  pattern recognition workshops}, pages 852--863, 2018.

\bibitem{uhlich2019mixed}
Stefan Uhlich, Lukas Mauch, Fabien Cardinaux, Kazuki Yoshiyama, Javier~Alonso
  Garcia, Stephen Tiedemann, Thomas Kemp, and Akira Nakamura.
\newblock Mixed precision dnns: All you need is a good parametrization.
\newblock {\em arXiv preprint arXiv:1905.11452}, 2019.

\bibitem{wan2020fbnetv2}
Alvin Wan, Xiaoliang Dai, Peizhao Zhang, Zijian He, Yuandong Tian, Saining Xie,
  Bichen Wu, Matthew Yu, Tao Xu, Kan Chen, et~al.
\newblock Fbnetv2: Differentiable neural architecture search for spatial and
  channel dimensions.
\newblock In {\em Proceedings of the IEEE/CVF Conference on Computer Vision and
  Pattern Recognition}, pages 12965--12974, 2020.

\bibitem{wang2018non}
Xiaolong Wang, Ross Girshick, Abhinav Gupta, and Kaiming He.
\newblock Non-local neural networks.
\newblock In {\em Proceedings of the IEEE conference on computer vision and
  pattern recognition}, pages 7794--7803, 2018.

\bibitem{wang2021knowledge}
Yiran Wang, Xingyi Li, Min Shi, Ke Xian, and Zhiguo Cao.
\newblock Knowledge distillation for fast and accurate monocular depth
  estimation on mobile devices.
\newblock In {\em Proceedings of the IEEE/CVF Conference on Computer Vision and
  Pattern Recognition Workshops}, pages 0--0, 2021.

\bibitem{wofk2019fastdepth}
Diana Wofk, Fangchang Ma, Tien-Ju Yang, Sertac Karaman, and Vivienne Sze.
\newblock Fastdepth: Fast monocular depth estimation on embedded systems.
\newblock In {\em 2019 International Conference on Robotics and Automation
  (ICRA)}, pages 6101--6108. IEEE, 2019.

\bibitem{wu2019fbnet}
Bichen Wu, Xiaoliang Dai, Peizhao Zhang, Yanghan Wang, Fei Sun, Yiming Wu,
  Yuandong Tian, Peter Vajda, Yangqing Jia, and Kurt Keutzer.
\newblock Fbnet: Hardware-aware efficient convnet design via differentiable
  neural architecture search.
\newblock In {\em Proceedings of the IEEE/CVF Conference on Computer Vision and
  Pattern Recognition}, pages 10734--10742, 2019.

\bibitem{xian2018monocular}
Ke Xian, Chunhua Shen, Zhiguo Cao, Hao Lu, Yang Xiao, Ruibo Li, and Zhenbo Luo.
\newblock Monocular relative depth perception with web stereo data supervision.
\newblock In {\em Proceedings of the IEEE Conference on Computer Vision and
  Pattern Recognition}, pages 311--320, 2018.

\bibitem{xie2017aggregated}
Saining Xie, Ross Girshick, Piotr Doll{\'a}r, Zhuowen Tu, and Kaiming He.
\newblock Aggregated residual transformations for deep neural networks.
\newblock In {\em Proceedings of the IEEE conference on computer vision and
  pattern recognition}, pages 1492--1500, 2017.

\bibitem{yang2019quantization}
Jiwei Yang, Xu Shen, Jun Xing, Xinmei Tian, Houqiang Li, Bing Deng, Jianqiang
  Huang, and Xian-sheng Hua.
\newblock Quantization networks.
\newblock In {\em Proceedings of the IEEE/CVF Conference on Computer Vision and
  Pattern Recognition}, pages 7308--7316, 2019.

\bibitem{zhang2021asimple}
Ziyu Zhang, Yicheng Wang, Zilong Huang, Guozhong Luo, Gang Yu, and Bin Fu.
\newblock A simple baseline for fast and accurate depth estimation on mobile
  devices.
\newblock In {\em Proceedings of the IEEE/CVF Conference on Computer Vision and
  Pattern Recognition Workshops}, pages 0--0, 2021.

\end{thebibliography}

}

\end{document}